\documentclass[twocolumn,floatfix,showpacs]{revtex4}
\usepackage{graphicx}
\usepackage{amsmath}
\usepackage{bm}
\begin{document}

\title{How to detect the pseudospin-$\bm{\frac{1}{2}}$ Berry phase in a photonic crystal with a Dirac spectrum}
\author{R. A. Sepkhanov}
\affiliation{Instituut-Lorentz, Universiteit Leiden, P.O. Box 9506, 2300 RA Leiden, The Netherlands}
\author{Johan Nilsson}
\affiliation{Instituut-Lorentz, Universiteit Leiden, P.O. Box 9506, 2300 RA Leiden, The Netherlands}
\author{C. W. J. Beenakker}
\affiliation{Instituut-Lorentz, Universiteit Leiden, P.O. Box 9506, 2300 RA Leiden, The Netherlands}
\date{May 2008}
\begin{abstract}
We propose a method to detect the geometric phase produced by the Dirac-type band structure of a triangular-lattice photonic crystal. The spectrum is known to have a conical singularity (= Dirac point) with a pair of nearly degenerate modes near that singularity described by a spin-$\frac{1}{2}$ degree of freedom (= pseudospin). The geometric Berry phase acquired upon rotation of the pseudospin is in general obscured by a large and unspecified dynamical phase. We use the analogy with graphene to show how complementary media can eliminate the dynamical phase. A transmission minimum results as a direct consequence of the geometric phase shift of $\pi$ acquired by rotation of the pseudospin over $360^{\circ}$ around a perpendicular axis. We support our analytical theory based on the Dirac equation by a numerical solution of the full Maxwell equations. 
\end{abstract}
\maketitle

\section{Introduction}
\label{intro}

Geometric phases (also known as Berry phases) typically appear in optics and quantum mechanics when a spin degree of freedom is transported along a closed orbit \cite{Sha89}. The geometric phase is given by the product of the enclosed solid angle and the spin, independently of the duration of the orbit (hence the adjective ``geometric''). 

The spin is usually $\frac{1}{2}$ in the quantum mechanical context, when the spin is the electron spin. In the optical context, the spin corresponds to the light polarization and may be either $\frac{1}{2}$ or 1 depending on whether the photon momentum is cycled or kept fixed \cite{Bha97}. An early experimental detection of the spin-$1$ geometric phase of a photon was the measurement of the rotating linear polarization in a twisted optical fiber \cite{Tom86}. For electrons, the recently observed \cite{Nov05,Zha05} anomalous quantization of Landau levels in graphene is a direct manifestation of the geometric phase of $\pi$ acquired by a pseudospin-$\frac{1}{2}$ which rotates over $360^{\circ}$ in a cyclotron orbit (since the pseudospin is tangential to the velocity).

The graphene example is unusual because the spin-$\frac{1}{2}$ that is rotating is not the true electron spin but an orbital degree of freedom with the same SU(2) symmetry, emerging from the motion of the electron in the periodic potential of the carbon atoms. Such a pseudospin is not tied to the fermionic statistics of the electrons and so it might also manifest itself in the bosonic optical context. 

The optical analogue of graphene is a photonic crystal with a two-dimensional (2D) triangular lattice structure. Haldane and Raghu \cite{Hal05} showed that a pair of almost degenerate Bloch waves $(\Psi_{1},\Psi_{2})\equiv\Psi$ near a $K$-point of the Brillouin zone can be represented by a pseudospin, coupled to the orbital motion. The wave equation,
\begin{subequations}
\label{Dirac}
\begin{align}
&H\Psi=\varepsilon\Psi,\;\;\varepsilon=\frac{\omega-\omega_{D}}{v_{D}},\label{Diraca}\\
&H=-i\sigma_{x}\frac{\partial}{\partial x}-i\sigma_{y}\frac{\partial}{\partial y}+\mu\sigma_{z},\label{Diracb}
\end{align}
\end{subequations}
is the 2D Dirac equation of a spin-$\frac{1}{2}$ particle with mass $\mu$ (nonzero if inversion symmetry is broken) \cite{Hal05}. The resulting dispersion relation,
\begin{equation}
\varepsilon^{2}=k_{x}^{2}+k_{y}^{2}+\mu^{2}, \label{dispersion}
\end{equation}
reduces to a double cone in the case $\mu=0$ of a perfect lattice, with a degeneracy at the frequency $\omega_{D}$ of the Dirac point. The slope $d\omega/dk=v_{D}$ is the frequency-independent group velocity. The upper cone (frequencies $\omega>\omega_{D}$) corresponds to the conduction band in graphene, and the lower cone ($\omega<\omega_{D}$) to the valence band. Several analogies between the electronic and optical transport properties near the Dirac point have been analysed \cite{Hal05,Sep07,Zha07,Gar07}. What is missing is an optical way to directly observe the geometric phase due to the rotating pseudospin, analogous to the ``smoking gun'' found in the electronic cyclotron motion \cite{Nov05,Zha05}.

A direct analogy is problematic because there exists no optical cyclotron motion. One can imagine other ways to have a photon execute a closed orbit, but the large and unspecified dynamical phase is likely to obscure the geometric phase. Here we show how complementary media \cite{Pen03} can be used to eliminate the dynamical phase, resulting in a transmission minimum that is a direct consequence of the $\pi$ phase shift acquired by the rotating pseudospin. We support our argument by an analytical solution of the Dirac equation and by a numerical solution of the full Maxwell equations.

\section{Calculation of the geometric phase}
\label{calculation}

\begin{figure}[tb]
\centerline{\includegraphics[width=1.0\linewidth]{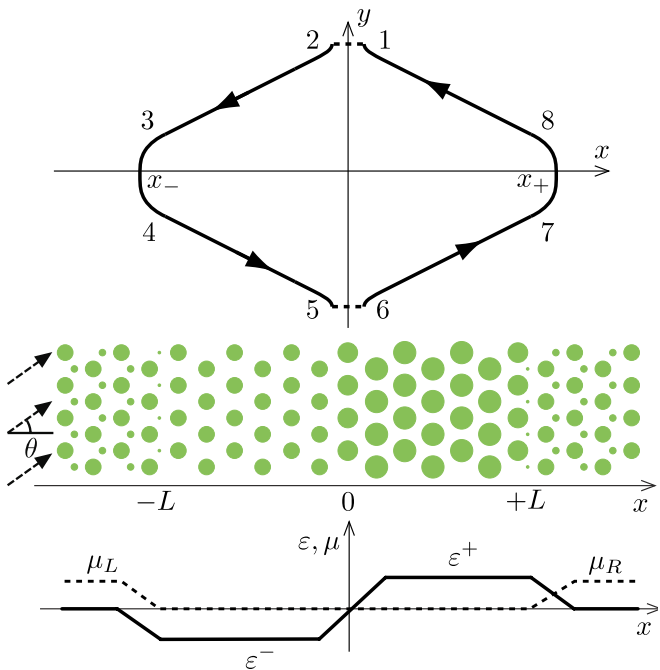}}
\caption{\label{fig_layout}
Lower two panels: Schematic layout (not to scale) of the triangular-lattice photonic crystal (with the cross-section of the dielectric rods shown in green) and plot of the corresponding profiles $\varepsilon(x)$ and $\mu(x)$. The two regions $x<0$ and $x>0$ form complementary media if the rescaled frequency $\varepsilon$ is an odd function of $x$ while the mass term $\mu$ is an even function of $x$. The top panel shows a closed orbit in the photonic crystal, with the dashed lines indicating tunneling through the region of imaginary wave vector.
}
\end{figure}

\begin{figure}[tb]
\centerline{\includegraphics[width=0.8\linewidth]{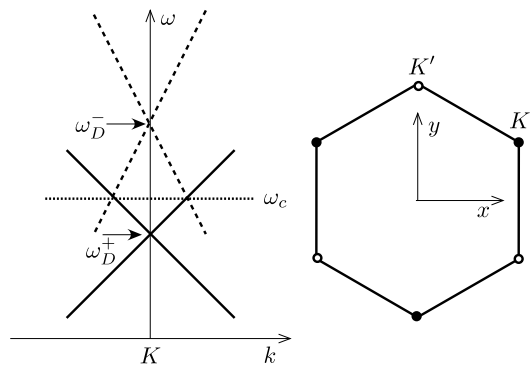}}
\caption{\label{fig_cones}
Left panel: Schematic view of the conical dispersion relations near a $K$-point, in the region $x>0$ (solid lines) and $x<0$ (dashed lines). The horizontal dotted line indicates the frequency $\omega_{c}$, given by Eq.\ \eqref{complementary_omega}, at which the two regions form complementary media. Right panel: Hexagonal first Brillouin zone of the triangular lattice. The $K$ and $K'$-points are indicated by filled and open dots, respectively. Only the $K$-points are excited in the geometry of Fig.\ \ref{fig_layout}.
}
\end{figure}

The system that can isolate the geometric phase from the dynamical phase is illustrated in Fig.\ \ref{fig_layout} (lower two panels). It is the optical analogue of the \textit{p-n} junction in graphene studied in Ref.\ \cite{Bee08}. In graphene, complementary media are formed when the Fermi level crosses from the conduction band to the valence band \cite{Che07}. For the optical analogue, we introduce a (smooth) step in the Dirac frequency at $x=0$, so that $\omega_{D}$ decreases from $\omega_{D}^{-}$ for $x<0$ to $\omega_{D}^{+}$ for $x>0$. The Dirac frequency can be changed for instance by varying the radius of the dielectric rods that form the photonic crystal. Unlike in the electronic case, a shift of $\omega_{D}$ is generally accompanied by a shift of $v_{D}$, from $v_{D}^{-}$ to $v_{D}^{+}$. The corresponding shift in the parameter $\varepsilon$ is from $\varepsilon^{-}$ to $\varepsilon^{+}$. We define the complementarity frequency $\omega_{c}$ such that
\begin{equation}
\frac{\omega_{c}-\omega_{D}^{+}}{v_{D}^{+}}= -\frac{\omega_{c}-\omega_{D}^{-}}{v_{D}^{-}}\Leftrightarrow\varepsilon^{+}=-\varepsilon^{-}.\label{complementary_omega}
\end{equation}
As illustrated in Fig.\ \ref{fig_cones}, waves of frequency $\omega_{c}$ have the same wave vector in absolute value in the two regions $x<0$ and $x>0$, but of opposite orientation relative to the group velocity (since $k$ and $d\omega/dk$ have the same sign for $x>0$ and opposite sign for $x<0$). Dynamical phase shifts accumulated in the two regions thus cancel, leaving only the geometric phase from the rotation of the pseudospin.

We calculate the geometric phase for the closed orbit shown in Fig.\ \ref{fig_layout} (top panel). Notice the negative refraction \cite{Che07,Not00} at the interface $x=0$ where the orbit tunnels between the upper and lower cones of the dispersion relation (Klein tunneling). The component $k_{y}=q$ of the wave vector parallel to the interface is conserved (because of translational invariance in the $y$-direction), while the component $k_{x}=k$ changes sign when $x\mapsto -x$. The orbit is reflected at the turning points $x_{\pm}$ by a mass term $\mu(x)$. We require $\mu(-x)=\mu(x)$ and $\varepsilon(-x)= -\varepsilon(x)$. Because $\Psi(x)$ and $\sigma_{x}\Psi(-x)$ are then both solutions of Eq.\ \eqref{Dirac} (for a given $y$-dependence $\propto e^{iqy}$), it follows that the transfer matrix $M(x,x')$ through the photonic crystal [defined by $\Psi(x)=M(x,x')\Psi(x')$] satisfies
\begin{equation}
M(x,0)\sigma_{x}M(0,-x)=\sigma_{x}.\label{Mrelation}
\end{equation}
This is a generalized complementarity relation~\cite{Kob06} (the original complementarity relation \cite{Pen03} would have the unit matrix in place of $\sigma_{x}$).

\begin{figure}[tb]
\centerline{\includegraphics[width=0.6\linewidth]{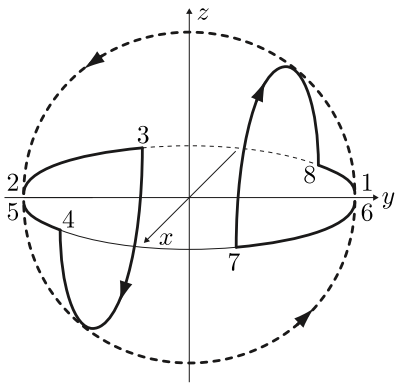}}
\caption{\label{fig_blochsphere}
Rotation of the Bloch vector $\bm{B}$ along the closed orbit of Fig.\ \ref{fig_layout}, with the corresponding points numbered. The full rotation sweeps out a solid angle of $2\pi$, producing a Berry phase of $\pi$.
}
\end{figure}

A trajectory description is applicable if the variations of $\mu$, $\omega_{D}$, and $v_{D}$ with $x$ are smooth on the scale of the wave length. The spatial derivatives in Eq.\ \eqref{Dirac} may then be replaced by the local wave vector, $-i\nabla\rightarrow\bm{k}$ (measured relative to the $K$ point). The solution is
\begin{equation}
\Psi={\cal C}^{-1/2}\begin{pmatrix}
\mu+\varepsilon\\
k+iq
\end{pmatrix}\equiv\begin{pmatrix}
\cos(\theta/2)\\
e^{i\phi}\sin(\theta/2)
\end{pmatrix}
,\label{Psisolution}
\end{equation}
with $k$ determined from $\varepsilon,\mu,q$ through Eq.\ \eqref{dispersion} and ${\cal C}=(\mu+\varepsilon)^{2}+|k+iq|^{2}$ a normalization constant. The angles $\phi,\theta$ define the Bloch vector $\bm{B}=(\cos\phi\sin\theta,\sin\phi\sin\theta,\cos\theta)$, representing the direction of the pseudospin on the Bloch sphere. The rotation of the Bloch vector along the closed orbit is indicated in Fig. \ref{fig_blochsphere}.

The geometric phase $\Phi=\Omega/2$ is one half the solid angle $\Omega$ subtended at the origin by the rotating Bloch vector \cite{Sha89}. We distinguish three contributions to $\Omega$, a contribution $\Omega_{-}$ from the trajectory in the lower cone of the dispersion relation ($x<0$), a contribution $\Omega_{+}$ from the trajectory in the upper cone ($x>0$), and a contribution $\Omega_{K}$ from Klein tunneling between the two cones (through the interface $x=0$, indicated by dashed lines). The Bloch vector that sweeps out $\Omega_{\pm}$ is given by $\bm{B}=(k,q,\mu)/\varepsilon$. It follows from $k(-x)=-k(x)$, $\mu(-x)=\mu(x)$, $\varepsilon(-x)=-\varepsilon(x)$ that $\Omega_{+}=-\Omega_{-}$, so the two contributions from the upper and lower cones cancel. 

The contribution from Klein tunneling between the points $\pm \delta x$ has imaginary $k=i\kappa$. The sign of $\kappa$ is positive when tunneling towards positive $x$ (from the lower cone to the upper cone) and negative when tunneling towards negative $x$ (from upper to lower cone) --- to ensure a decaying wave $\propto e^{-\kappa x}$. 

The Bloch vector
\begin{equation}
\bm{B}=(\mu^{2}+q^{2})^{-1}\begin{pmatrix}
0\\
\kappa\mu+q\varepsilon\\
\mu\varepsilon-\kappa q
\end{pmatrix}\label{BlochKlein}
\end{equation}
rotates in the $y-z$ plane from $\bm{B}_{+}$ to $\bm{B}_{-}$ through the positive $z$-axis (tunneling from upper to lower cone) and back to $\bm{B}_{+}$ through the negative $z$-axis (tunneling from lower to upper cone). The value of $\bm{B}_{\pm}$ of the Bloch vector at points $\pm\delta x$ follows from Eq.\ \eqref{BlochKlein} with $\kappa=0$,
\begin{equation}
\bm{B}_{\pm}=\frac{1}{\varepsilon(\pm\delta x)}\begin{pmatrix}
0\\
q\\
\mu(\pm\delta x)\end{pmatrix}
\Rightarrow\bm{B}_{-}=-\bm{B}_{+}.\label{Bpm}
\end{equation}
The resulting $360^{\circ}$ rotation of $\bm{B}$ in the $y-z$ plane sweeps out a solid angle $\Omega_{K}=2\pi$, so that the total geometric phase acquired in the closed orbit of Fig.\ \ref{fig_layout} is $\Phi=\pi$.

\section{Destructive interference of partial waves}
\label{destructive}

The Berry phase of $\pi$ suppresses the formation of a bound state at the complementarity frequency $\omega_{c}$. To show this, we demonstrate the destructive interference of partial waves that return to the point of origin after multiple tunnel events. A more formal proof of the absence of a bound state at $\omega_{c}$ is given in App.\ \ref{noboundstate}.

\begin{figure}[tb]
\centerline{\includegraphics[width=1.0\linewidth]{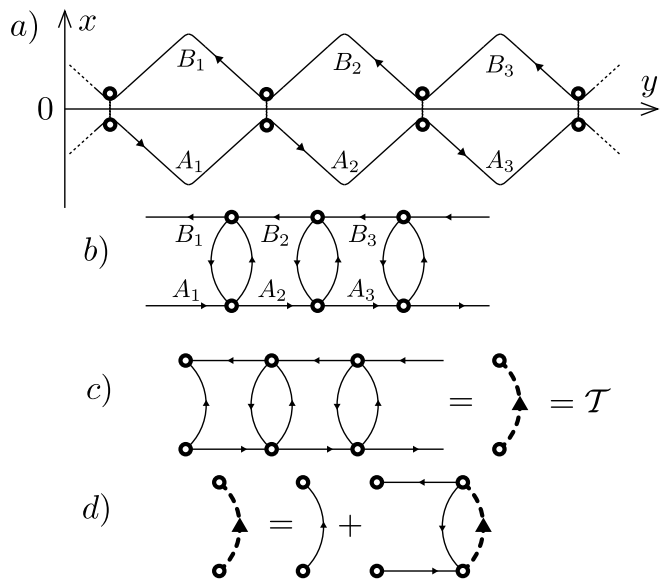}}
\caption{
Sequence of partial wave amplitudes $A_{n}$ and $B_{n}$, produced by tunnel events (black circles) at the interface $x=0$ between two complementary media. Panel a) shows a top view of the multiply scattered rays, panel b) shows a more abstract representation. Panels c) and d) illustrate the construction of the total transmission amplitude ${\cal T}$ and of the Dyson equation that it obeys.
}
\label{fig:summation}
\end{figure}

The scattering problem is illustrated in Fig.\ \ref{fig:summation}. Partial wave amplitudes are labelled $A_{n}$ for $x<0$ and $B_{n}$ for $x>0$. The wave amplitudes at a tunnel event (black circle) are related by a unitary scattering matrix,
\begin{equation}
\begin{pmatrix}
A_{n+1}\\
B_{n}
\end{pmatrix}=S
\begin{pmatrix}
A_{n}\\
B_{n+1}
\end{pmatrix},\;\;S=\begin{pmatrix}
r&t'\\
t&r'
\end{pmatrix}.\label{SABdef}
\end{equation}
The phase shift of $\pi$ acquired in a single closed loop $A_{n}\rightarrow B_{n}\rightarrow A_{n}$ implies 
\begin{equation}
{\rm arg}\,(t)+{\rm arg}(t')=\pi\Rightarrow
t'=-t^{\ast}.\label{Berryphase}
\end{equation}
Unitarity of $S$ then requires that the scattering matrix of a tunnel event is of the form
\begin{equation}
S=\begin{pmatrix}
r&-t^{\ast}\\
t&r^{\ast}
\end{pmatrix},\;\;|r|^{2}+|t|^{2}=1.\label{Sconstrained}
\end{equation}

An initial wave amplitude $A_{n}^{\rm initial}$ interferes with the sum $A_{n}^{\rm final}$ of partial wave amplitudes that return after different sequences of tunnel events. Each sequence $A_{n}\rightarrow\cdots\rightarrow B_{n}\rightarrow\cdots\rightarrow A_{n}$ includes $B_{n}$ exactly once. We write $A_{n}^{\rm final}={\cal T}'{\cal T}A_{n}^{\rm initial}$, with ${\cal T}$ the total transmission amplitude from $A_{n}^{\rm initial}$ to $B_n$ and ${\cal T}'$ the total transmission amplitude from $B_{n}$ to $A_{n}^{\rm final}$. 

For ${\cal T}$ we can construct a Dyson equation (see Fig.\ \ref{fig:summation}):
\begin{align}
{\cal T} &= t + r' {\cal T}r + r'{\cal T}t' {\cal T}r 
+ r'{\cal T} ( t' {\cal T} )^2r  + \cdots\\
&=t + \frac{rr'{\cal T}}{1 - t'{\cal T}}.
\label{TDyson}
\end{align}
Similarly, we have
\begin{equation}
{\cal T}'=t'+\frac{rr'{\cal T}'}{1 - t{\cal T}'}.\label{TprimeDyson}
\end{equation}
The two Dyson equations can be combined into a single equation for the variable $\xi={\cal T}/t={\cal T}'/t'$,
\begin{equation}
\xi=1+\frac{rr'\xi}{1-tt'\xi}.\label{xiDyson}
\end{equation}

At this point we invoke the Berry phase relation \eqref{Berryphase}, which together with unitarity implies $tt'=-|t|^{2}=rr'-1$. The Dyson equation \eqref{xiDyson} then reduces to
\begin{equation}
\xi^{2}=1/|t|^{2}.\label{xiresult}
\end{equation}
Regardless of the ambiguity in the sign of $\xi$, we can conclude that
\begin{equation}
{\cal T}'{\cal T}\equiv\xi^{2}t't=-1\Rightarrow A_{n}^{\rm final}=-A_{n}^{\rm initial}.\label{TTprimeresult}
\end{equation}
The end result is therefore a phase shift of $\pi$ between $A_{n}^{\rm final}$ and $A_{n}^{\rm initial}$, without any change in the magnitude. The destructive interference of $A_{n}^{\rm final}$ and $A_{n}^{\rm initial}$, which prevents the formation of a bound state at frequency $\omega_{c}$, is a direct consequence of the phase shift of $\pi$ acquired in a single closed loop, even if the weight $|t|^{2}$ of a single loop is small.

\section{Detection of the destructive interference}
\label{detection}

\begin{figure}[tb]
\centerline{\includegraphics[width=0.8\linewidth]{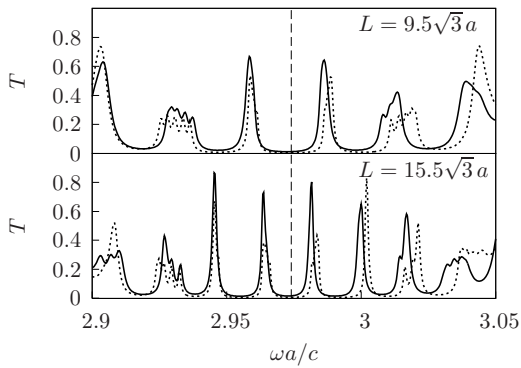}}
\caption{\label{mass_even}
Frequency dependent transmission probability for two values of $L$ in the case $\mu_R = \mu_L$ of complementary media. The solid curves show the numerical result from the Maxwell equations, while the dashed curves are calculated analytically from the Dirac equation. The vertical dashed line indicates the complementarity frequency $\omega_{c}$.
}
\end{figure}

\begin{figure}[tb]
\centerline{\includegraphics[width=0.8\linewidth]{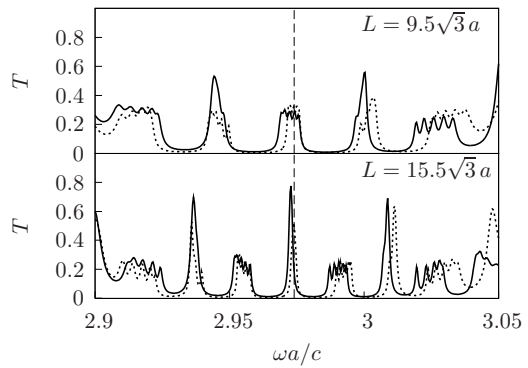}}
\caption{\label{mass_odd}
The same as Fig.\ \ref{mass_even}, for the case $\mu_R = -\mu_L$ when the complementarity is broken by the mass term.
}
\end{figure}

To detect the destructive interference, we propose a measurement of the transmission probability $T$ of resonant tunneling of a plane wave through the photonic crystal \cite{note1}. If the confinement at $x=\pm L$ is strong, the transmission probability will have narrow resonances at the frequencies of the quasi-bound states. The destructive interference at $\omega=\omega_{c}$ will produce a transmission minimum for any $L$. This is unlike usual Fabry-Perot resonances, which would shift with $L$, so that there would not be a systematic minimum or maximum at any particular frequency. 

For a well-developed conical band structure we take the TE polarization (magnetic field parallel to the dielectric rods). The parameters of the photonic crystal are summarized in a footnote \cite{note2}. A 7\% increment of the radius of the rods (at fixed lattice constant $a$) shifts $\omega_{D}$ and $v_{D}$ by about 5\% and 15\%, respectively. The mass term at $x=\pm L$ is created by breaking the inversion symmetry through the addition of an extra rod in the unit cell (see Fig.\ \ref{fig_layout}). We have solved the full Maxwell equations with the finite-difference time-domain method \cite{Taf05} using the {\sc meep} software package \cite{Far06}. The wave vector $\bm{k}=(k,q)$ is the displacement of the wave vector of an incident plane wave from the $K$-point at wave vector $\bm{K}=\frac{2}{3}\pi a^{-1}(\sqrt{3}, 1)$. There are two inequivalent $K$-points in the hexagonal first Brillouin zone, see Fig.\ \ref{fig_cones}, and we excite a single one by orienting the lattice relative to the incident plane wave as indicated in Fig.\ \ref{fig_layout}. [The angle of incidence is spread over a narrow interval $\delta\theta\simeq 2.3^{\circ}$ around $\theta=\arcsin (cK_{y}/\omega_{c})$.] Results are shown in Fig.\ \ref{mass_even} (solid curves) for two values of $L$.

As an independent test on the accuracy of the numerical calculations, we have also calculated analytically the transmission probability from the Dirac equation \eqref{Dirac}, using the transfer matrix method of Ref.\ \cite{Sep07}. For simplicity we assumed in this analytical calculation an ideal coupling between the plane waves in free space and the Bloch waves in the photonic crystal. The analytical results are also plotted in Fig.\ \ref{mass_even} (dotted curves) and are found to agree well with the numerical results from the Maxwell equations. There are no adjustable parameters in this comparison. (The parameters $\omega_{D},v_{D},\mu$ were extracted independently from the band structure, calculated using the {\sc mpb} software package \cite{Joh01}.)

We observe in Fig.\ \ref{mass_even} a transmission minimum at $\omega_{c}$ that does not shift with variations of $L$. To test our interpretation of the origin of this minimum, we have broken the complementarity of the media by inverting the sign of the mass term at the left end of the crystal. (This can be done by inverting the position of the extra rod in the unit cell.) For $q=0$ the inversion produces an extra phase shift of $\pi$ that switches the destructive interference to constructive interference --- in agreement with the observed switch (see Fig.\ \ref{mass_odd}) from a transmission minimum to a transmission maximum at $\omega_{c}$.

\section{Conclusion}
\label{conclude}

In conclusion, we have proposed a method to detect the pseudospin-$\frac{1}{2}$ geometric phase produced by the Dirac spectrum in a photonic crystal. The dynamical phase can be eliminated by measuring the transmission through complementary media, so that only the $\pi$ geometric phase remains and a parameter-independent transmission minimum results at the complementarity frequency. Our analysis is based on the Dirac equation, which is an approximate long-wave length description, but it is fully supported by an exact numerical solution of the Maxwell equations in a triangular lattice of dielectric rods.

The experiment proposed and analysed here can be seen as the optical analogue of the detection of the geometric phase acquired during electronic cyclotron motion in graphene \cite{Nov05,Zha05}. There is one fundamental difference: In a cyclotron orbit the $\pi$ phase shift is produced by $360^{\circ}$ rotation of the pseudospin in the $x-y$ plane of the lattice, while in our complementary media the rotation is in the perpendicular $y-z$ plane. The difference shows up in the dependence of the geometric phase on a mass term $\mu\sigma_{z}$ in the Dirac equation. A nonzero mass pushes the pseudospin out of the $x-y$ plane, thereby reducing the enclosed solid angle and hence reducing the geometric phase acquired during a cyclotron orbit \cite{Car08}. In the complementary media the geometric phase remains equal to $\pi$.

In graphene, the suppression of the density of states at a \textit{p-n} junction is analogous to the proximity effect in a normal-superconductor junction \cite{Bee08}. Observation of the optical counterpart presented in this paper would open up the possibility to study superconducting analogies in nonelectronic systems.

\acknowledgments

We acknowledge discussions with A. R. Akhmerov, J. H. Bardarson, and M. J. A. de Dood. This research was supported by the Dutch Science Foundation NWO/FOM.

\appendix
\section{Absence of a bound state at the complementarity frequency}
\label{noboundstate}

The demonstration of destructive interference of partial waves given in Sec.\ \ref{destructive} explicitly shows how the Berry phase of $\pi$ prevents the formation of a bound state at the complementarity frequency $\omega_{c}$. A more formal proof, that does not rely on the partial wave decomposition, is given here.

We use again the property that if $\Psi(x)e^{iqy}$ is a solution of Eq.\ \eqref{Dirac} at $\omega=\omega_{c}$, then also $\sigma_{x}\Psi(-x)e^{iqy}$ is a solution at the same frequency. We may therefore take even and odd superpositions of these two states to form new bound states $\Psi_{\pm}$ that satisfy $\sigma_{x}\Psi_{\pm}(0)=\pm\Psi_{\pm}(0)$. The photon flux density through the interface $x=0$ is
\begin{equation}
v_{D}\Psi_{\pm}^{\ast}(0)\sigma_{x}\Psi_{\pm}(0)=\pm v_{D}|\Psi_{\pm}(0)|^{2}. \label{photonflux}
\end{equation}
This should vanish for a bound state, which is only possible if $\Psi_{\pm}(0)=0$, meaning that the two regions $x<0$ and $x>0$ are decoupled. Any tunnel coupling between the two regions will result in $\Psi_{\pm}(0)\neq 0$, preventing the formation of a bound state at $\omega_{c}$.


\begin{thebibliography}{99}
\bibitem{note2} The parallel dielectric rods in air that form the triangular-lattice photonic crystal (lattice constant $a$, dielectric constant $14$) have radius $0.268 \, a$ for $-L<x<0$ and $0.288 \, a$ for $0<x<L$. (We smoothly changed the radius between these two values over an interval of order $5\,a$.) The resulting parameters of the Dirac spectrum are $\omega_D^{-}=3.05\, c / a$, $v_D^{-}=0.395\, c$, $\omega_D^{+}=2.91\, c / a$, $v_D^{+}=0.342\, c$. The complementarity frequency is $\omega_{c} = 2.974 \, c / a$. The inversion symmetry is broken in the region $L<|x|<L+10\,a$ by inserting thin rods (radius $0.08\,a$) in the unit cell, displaced by $ \pm a(1/\sqrt{3},0)$ from the center of the thick rods (radius $0.271\,a$). The dispersion relation in that region has parameters $\omega_D=2.973\, c / a$, $v_D=0.313\, c$, $\mu=\pm 0.16/a$.
\end{thebibliography}
\end{document}